\documentclass[runningheads]{llncs}
\usepackage{graphicx}
\usepackage{amsmath,amssymb}
\usepackage{floatrow}
\usepackage{subcaption}
\usepackage{wrapfig}
\newfloatcommand{capbtabbox}{table}[][\FBwidth]

\usepackage{blindtext}
\begin{document}
\title{Information Bottleneck Attribution for Visual Explanations of Diagnosis and Prognosis}
\author{
Ugur Demir\inst{1} \thanks{Corresponding author (ugurdemir2023@u.northwestern.edu)} \and
Ismail Irmakci \inst{1,2} \and 
Elif Keles\inst{1} \and
Ahmet Topcu\inst{3} \and
Ziyue Xu\inst{4} \and
Concetto Spampinato\inst{5} \and
Sachin Jambawalikar\inst{6} \and
Evrim Turkbey\inst{7} \and
Baris Turkbey\inst{7} \and
Ulas Bagci\inst{1}
}

\institute{
Department of Radiology and BME, Northwestern University, Chicago, IL \and
ECE, Ege University, Izmir, Turkey \and
Tokat State Hospital, Tokat, Turkey \and
NVIDIA, Bethesda, MD \and
University of Catania, Catania, Italy \and
Columbia University Medical Center, New York, NY \and
National Cancer Institute, National Institutes of Health, Bethesda, MD
}
\maketitle              
\begin{abstract}
Visual explanation methods have an important role in the prognosis of the patients where the annotated data is limited or unavailable. There have been several attempts to use gradient-based attribution methods to localize pathology from medical scans without using segmentation labels. This research direction has been impeded by the lack of robustness and reliability. These methods are highly sensitive to the network parameters. In this study, we introduce a robust visual explanation method to address this problem for medical applications. We provide an innovative visual explanation algorithm for general purpose and as an example application we demonstrate its effectiveness for quantifying lesions in the lungs caused by the Covid-19 with high accuracy and robustness without using dense segmentation labels. 
This approach overcomes the drawbacks of commonly used Grad-CAM and its extended versions. The premise behind our proposed strategy is that the information flow is minimized while ensuring the classifier prediction stays similar. Our findings indicate that the bottleneck condition provides a more stable  severity estimation than the similar attribution methods. The source code will be publicly available upon publication.

\end{abstract}
\section{Introduction}
The role of visual explanation methods in deep learning received increased attention especially in high-risk applications.
Determining the impacts of each pixel in the input on the classifier decision has huge importance to understand the behavior of the classifiers. Results from earlier studies demonstrate that these methods can be used to estimate coarse \textit{heatmaps} (responses) of a deep network to a specific input \cite{Florian2020Weakly}. Grad-CAM \cite{SelvarajuGradCAM2017} and its variants have become a dominant approach for visual explanation and they have been used to highlight areas that provide evidence in favor of, and against choosing a certain class. Furthermore, locating objects and even weakly segmenting the object of interest are some of the other applications that these methods have been applied to \cite{Florian2020Weakly,Li_2018_CVPR}. 
Despite its practicality and easy-to-use nature, Grad-CAM has significant drawbacks of (i) tuning its parameters, (ii) high number of false positives, and (iii) overestimating the response regions \cite{Eitel2019Testing,Young2019Deep}. Hence, the adoption of Grad-CAM and its variants is not completed and more trustable alternatives are needed to be developed, particularly for high-risk domains such as medicine. In this study, we develop a new visual attribution method that is superior to the commonly used Grad-CAM method. Although the proposed method is generic and can be applied to any classification-based applications, we demonstrate its efficacy in two important tasks in the medical imaging domain: diagnosis and prognosis. Due to the emergent need for helping the healthcare system to combat Covid-19, we focus on a Covid-19 problem where computed tomography scans are used for patient management. 

\textbf{Covid-19 Diagnosis and Prognosis} When the pandemic started, the researchers put lots of effort to develop automated diagnostic systems to fight against Covid-19. The main problem was the lack of knowledge about the disease and there was not enough resource to annotate dense and large datasets. 
Despite rapid and reliable advances in diagnostic tool, determining the severity of the Covid-19 positive patient remains a challenging goal, a crucial step for hospitalization planning and after Covid-10 clinics. Segmentation can be used to find morphometry and volumetry of lesions caused by Covid-19 but this requires labor-intensive segmentation labels from expert radiologists. More recent studies start to address the problem of prognosis  (severity estimation) of Covid-19 patients from CT scans and Grad-CAM is still the most used visual explanation method therein \cite{HarmonArtificial2020}. Lesions caused by Covid-19 have been identified by Grad-CAM without using segmentation ground truths \cite{HarmonArtificial2020,HarshAdeep2020}. Since Grad-CAM-based approaches are not robust, they overestimate the related regions in the images, and often difficult to tune its parameters to find reliable visualization, there is a strong need to develop an alternative visualization method addressing the current drawbacks of the Grad-CAM.

\textbf{Information Bottleneck} There is an increase in visual explanation methods that generate the heatmap in a more constrained way to ensure it focuses on the relevant region. In \cite{schulz2020iba}, a new method, Information Bottleneck Attribution (IBA), is introduced for an alternative visual explanation method based on the information bottleneck concept. Basically, IBA injects a noise to a specified intermediate layer of a network to mask out uninformative regions. The noise matrix is learned by an optimization process that tries to minimize the mutual information between the original features and the masked features. To prevent the trivial solution, it enforces the network to keep its prediction constant. This optimization promises to learn the mask that keeps the minimal enough information necessary for the prediction. 
Inspired by the IBA, 
we 
propose a conceptual theoretical framework based on information bottleneck that can both predict Covid-19 presence and estimate its severity from CT scans. We show that IBA results are more robust and reliable than the Grad-CAM. The visual results are inspected by four expert physicians to measure the quality of pathology localization as well as comparisons between the proposed IBA and the baseline Grad-CAM. Our experiments show that the results align with the lesions without having ground truth masks. 


\begin{figure}
\centering
\includegraphics[width=0.9\textwidth]{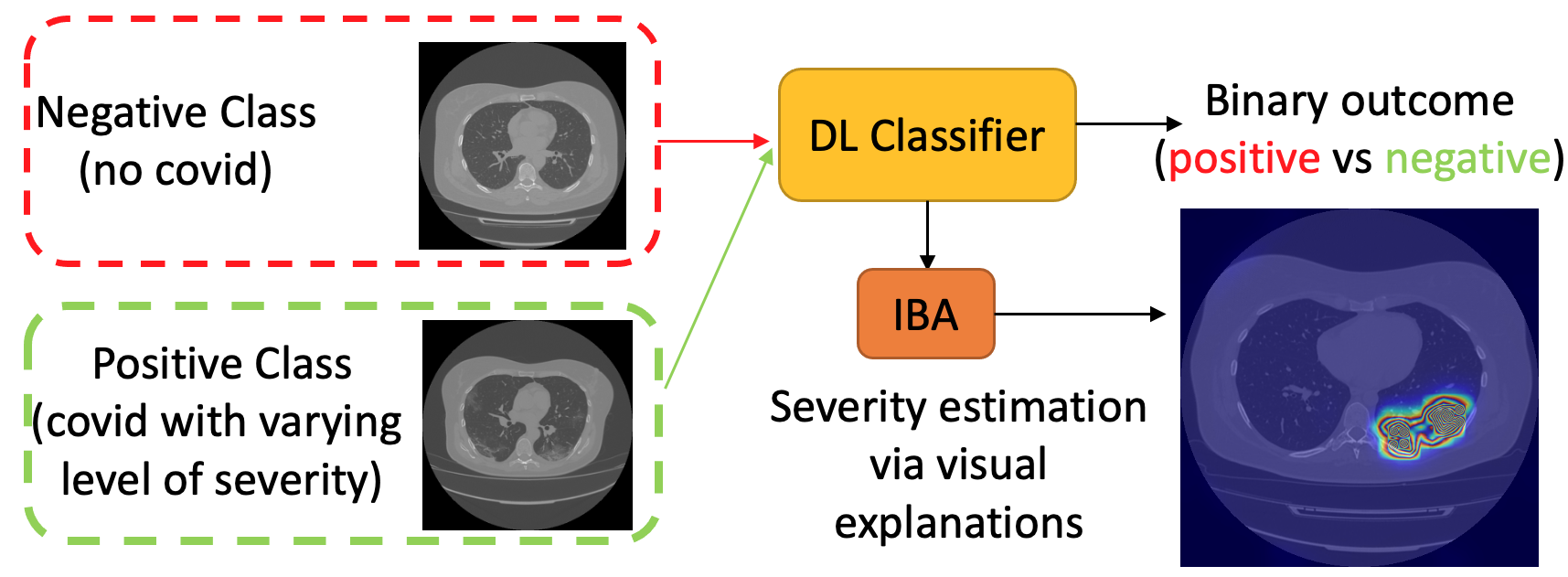}
\caption{Overview of information bottleneck based weak lesion localization. A binary classifier is trained independently with the positive and negative samples.} \label{fig:overview}
\end{figure}

\section{Related Work}
\textbf{Visual Explanation Methods} There has been a significant number of studies that focus on the CNN visualization \cite{SelvarajuGradCAM2017,Chattopadhay2018Gradcamplus,schulz2020iba}. 
In \cite{SelvarajuGradCAM2017}, convolutional features and their gradient are used to estimate an importance map for the input image. These heatmaps indicate image regions that have discriminative features. However, the gradients can be unstable for certain inputs that will result in irrelevant outputs. In \cite{schulz2020iba}, information bottleneck \cite{TishbyInformation1999} approach was utilized for visual explanation. It learns a mask that controls the information flow in the intermediate layers of a neural network to filter out irrelevant regions. It provides a theoretical guarantee that the masked out regions are unnecessary for the prediction; this 
is crucial for more robust visual explanation results.

\begin{figure}
\centering
\includegraphics[width=0.9\textwidth]{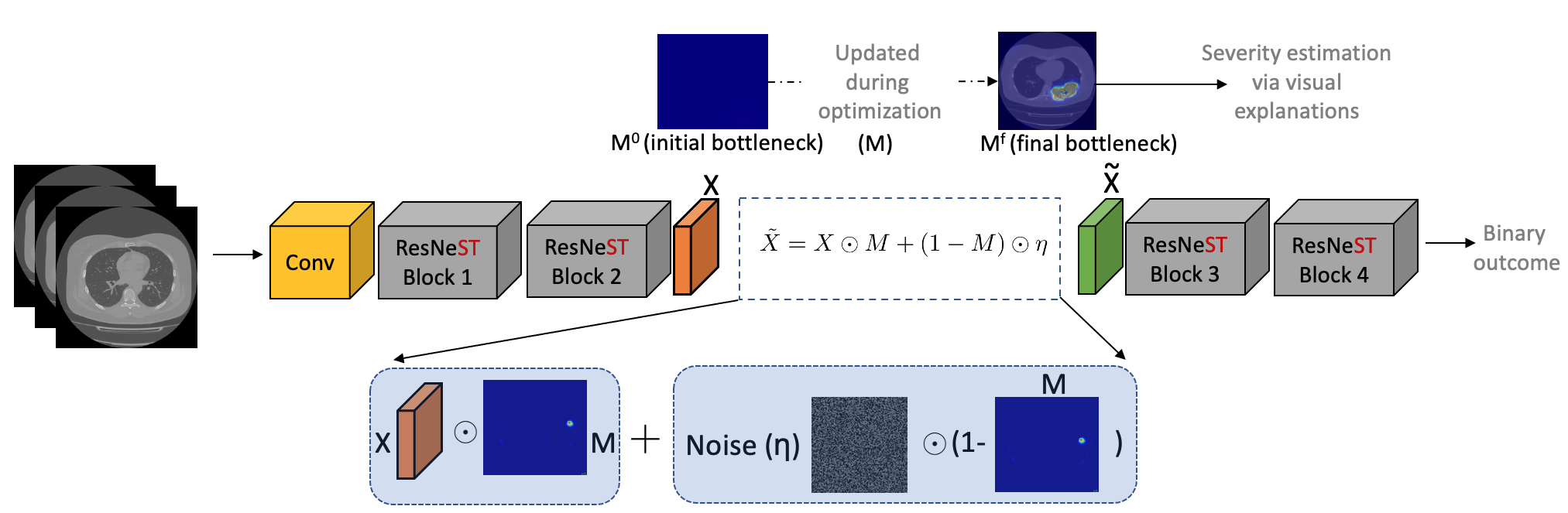}
\caption{Overall architecture. IBA method is applied to the pretrained ResNeSt.} 
\label{fig:resnest_iba}
\end{figure}

\textbf{Covid-19 Severity Estimation} 
Instead of diagnostic applications, recent studies focus on the severity of the COVID-19 cases where the spread of the lesions is estimated \cite{HarmonArtificial2020,ZekunAnovel2021,TangSeverity2021,GuillaumeAI2021,LiCT2020,HarshAdeep2020,ShanAbnormal2020}. In \cite{TangSeverity2021}, lungs are segmented and variety of features are extracted from the CT scans  to predict the severity of the patient. Well-curated pipelines for feature extraction and segmentation were challenges. Using Grad-CAM for visual explanation on Covid-19 was also another popular approach to localize important regions \cite{HarmonArtificial2020,HarshAdeep2020}. However, it is known that Grad-CAM-based approaches are sensitive to many parameters in the network, and they are not robust Here, we introduce a better visual explanation method to localize Covid-19 related lesions. To the best of our knowledge, this is the first study where Information Bottleneck based attribution is adopted for both diagnosis and prognosis tasks.


\section{Method}

We used Information Bottleneck based approach that  finds the critical input regions for the decision-making process. Empirical results show that those important regions align with Covid-19 related lesions. During the training, input scans are only considered as positive and negative samples instead of well and full labels. Figure \ref{fig:overview} shows the general structure of our approach. We utilize recently introduced ResNeSt architecture \cite{zhang2020resnest} as a binary classifier on 2D slices since it uses channel-wise attention and multi-paths to increase the representation capacity.  Other CNN architectures can be used as backbones instead of ResNeSt, we do not impose any restriction on the choice of networks.
We employ IBA to the trained binary Covid-19 classifier to obtain the heatmaps as shown in Figure \ref{fig:resnest_iba}. 


\subsection{Information Bottleneck Attribution}
Information bottleneck is a technique that compresses the information provided by the input associated with the label \cite{TishbyInformation1999}. Let $X$ and $Y$ be random variables associated with the input of the network and the prediction (output), respectively. In the standard classification setup, $p(y|x)$, the network uses the whole information from input X to predict Y. Information bottleneck introduces another random variable $\tilde{X}$ that is obtained by compressing the $X$. The $\tilde{X}$ is calculated by an optimization process that forces to use of only relevant information from the input related to $Y$. 
To find the optimal solution, the mutual information between $X$ and $\tilde{X}$ is minimized while the mutual information between $\tilde{X}$ and $Y$ is maximized for the $p(\tilde{X}|X)$;
\begin{equation}\label{eq:info_bottleneck}
\min_{p(\tilde{X}|X)} I(X;\tilde{X}) - \beta I(\tilde{X};Y)
\end{equation}
where $I(\cdot;\cdot)$ is the mutual information and $beta$ is the Lagrange multiplier that controls the amount of relevant information to flow.


In \cite{schulz2020iba}, information bottleneck is placed into a pre-trained network’s intermediate layer that still has spatial information. Information flow is reduced by injecting noise into the representation. Similar to that, $X$ and $\tilde{X}$ in our study express the original and the distorted representation, respectively. A matrix $M$ that has the same shape with $X$ controls where the noise will be injected spatially. The calculation is operated by
\begin{equation} \label{eq:noise_inject}
\tilde{X} = X \odot M + (1 - M) \odot \eta
\end{equation}
where $\odot$ is Hadamard product and $\eta$ is a random noise drawn from a certain distribution. 
The random noise is sampled from a normal distribution parametrized by the statistics of the original representation $X$; $\eta \sim \mathcal{N}(\mu_X,\,\sigma^{2}_X)$. 
For the Information Bottleneck optimization, we minimize the mutual information between $X$ and $\tilde{X}$  by
\begin{equation} \label{eq:mi_org}
I(X;\tilde{X}) = \mathbb{E}_X [D_{KL}(P_{\tilde{X}|X} || P_{\tilde{X}} )]
\end{equation}
where $P$ is a probability distribution and $D_{KL}$ is Kullback–Leibler divergence. Since calculating $P_{\tilde{X}}$ is intractable, it is replaced with the variation approximation $Q_Z \sim \mathcal{N}(\mu_X,\,\sigma^{2}_X)$. Then Equation \ref{eq:mi_org} becomes 
\begin{equation} \label{eq:mi_var}
I(X;\tilde{X}) = \mathbb{E}_X [D_{KL}(P_{\tilde{X}|X} || Q_{\tilde{X}} )] - D_{KL}(P_{\tilde{X}} || Q_{\tilde{X}}).
\end{equation}

In Equation \ref{eq:mi_var}, calculating the second is not trivial because of $P_{\tilde{X}}$. Removing it can cause overestimation of the actual mutual information,
\begin{equation}
\tilde{I}(X;\tilde{X}) = \mathbb{E}_X [D_{KL}(P_{\tilde{X}|X} || Q_{\tilde{X}} )],
\end{equation}
but this will still guarantee that if the value calculated by $\tilde{I}(X;\tilde{X})$ is zero, the information from that region will not be necessary for the prediction. Finally, if we put $\tilde{I}(X;\tilde{X})$ into Equation \ref{eq:info_bottleneck}, we can optimize the $M$ to find out important regions. We resized $M$ to the input size to use it as a visual explanation. 





\section{Datasets}
 
\begin{wraptable}{r}{0.5\linewidth}
\begin{tabular}{|c|c|c|c|}
\hline
Severity & Class & GGO & N \\
  \hline
  Zero & CT-0 & None (Healthy) & 254 \\
  Mild & CT-1 &  $\le$ 25\% & 684 \\
  Moderate & CT-2 & 25 - 50\% & 125 \\
  Severe & CT-3 & 50 - 75\% & 45 \\
  Critical & CT-4 & $\ge$ 75\% & 2 \\
  \hline
\end{tabular}
\caption{Ground glass opacity (GGO) distributions in each severity category with class labels and number of samples}
\label{table:data_class}
\end{wraptable}
We used MOSMEDDATA \cite{morozov2020mosmeddata} that contains lung CT scans from Covid-19 positive cases along with healthy lung scans. The dataset was collected in 2020 between March and April from municipal hospitals in Moscow, Russia. There were 1110 studies aged from 18 to 97 years old. It has 42\% male, 56\% female, and 2\% other/unknown subjects. MOSMEDDATA contains subjects from 5 categories from healthy to the critical level. Table \ref{table:data_class} shows the definition of categories and the number of samples per each class. During the training, we re-categorized the data as healthy and unhealthy for our binary classification task. The dataset was then split into training (70\%) and testing (30\%) sets without changing the class distribution. Also, we used lung segmentation maps to mask attribution results for both Grad-CAM and IBA approach.



\begin{figure}
    \includegraphics[height=3.90cm]{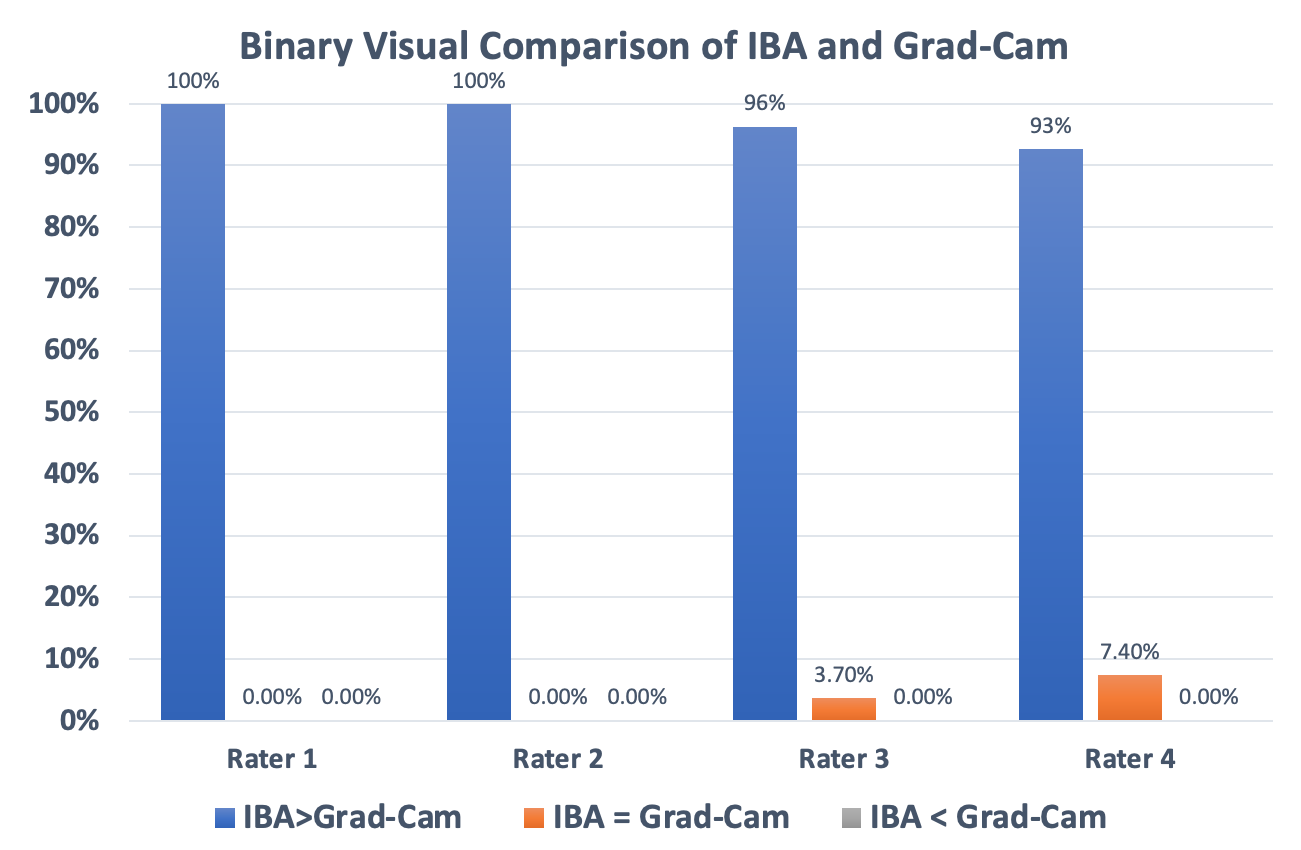}
    \includegraphics[height=3.91cm]{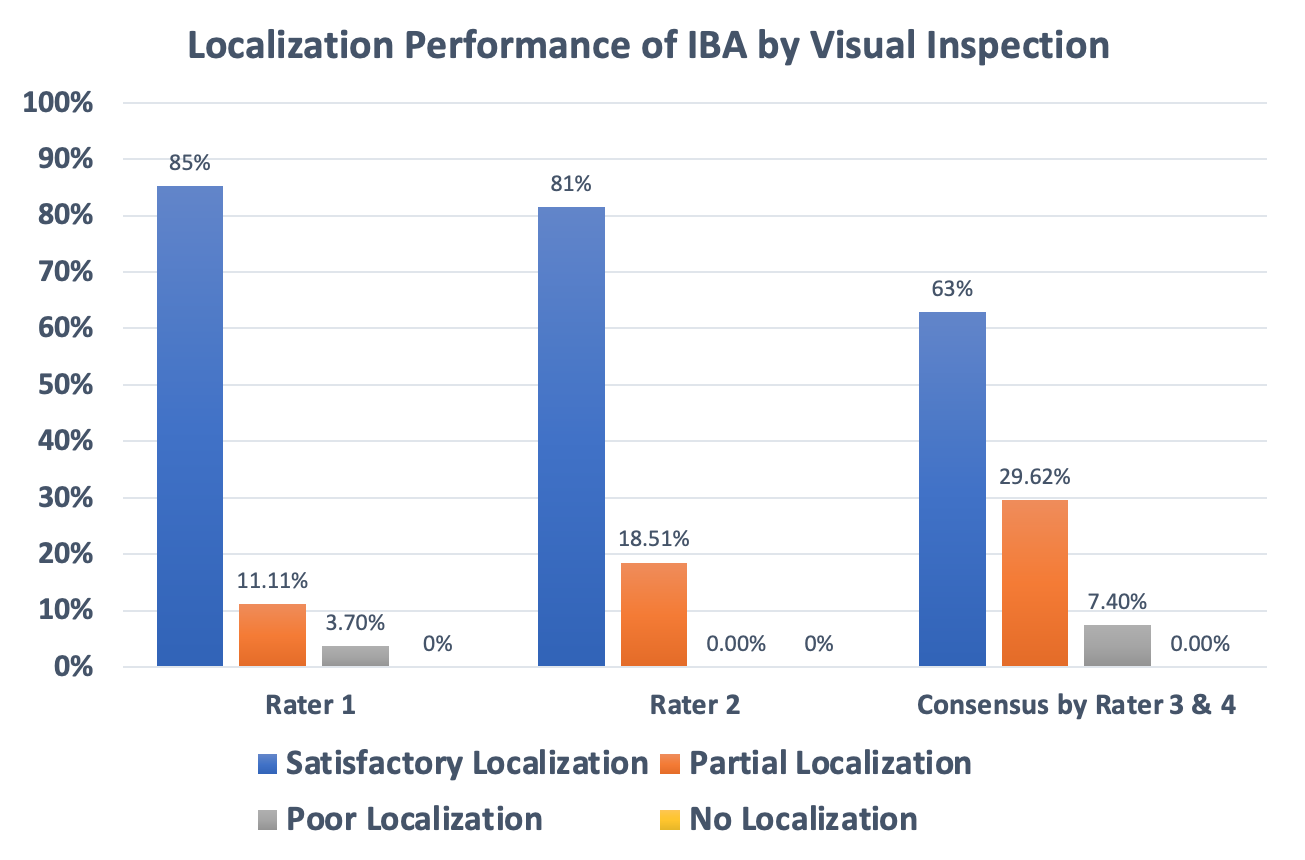}
    \caption{Visual Evaluation results. Left: Visual inspection scores for method comparison. Right: localization performance based based on the following scores: (1) satisfactory localization, (2) partial localization, (3) poor localization, and (4) no localization/unacceptable.}
    \label{fig:vis}
\end{figure}

\begin{figure}
\centering
\includegraphics[width=\textwidth]{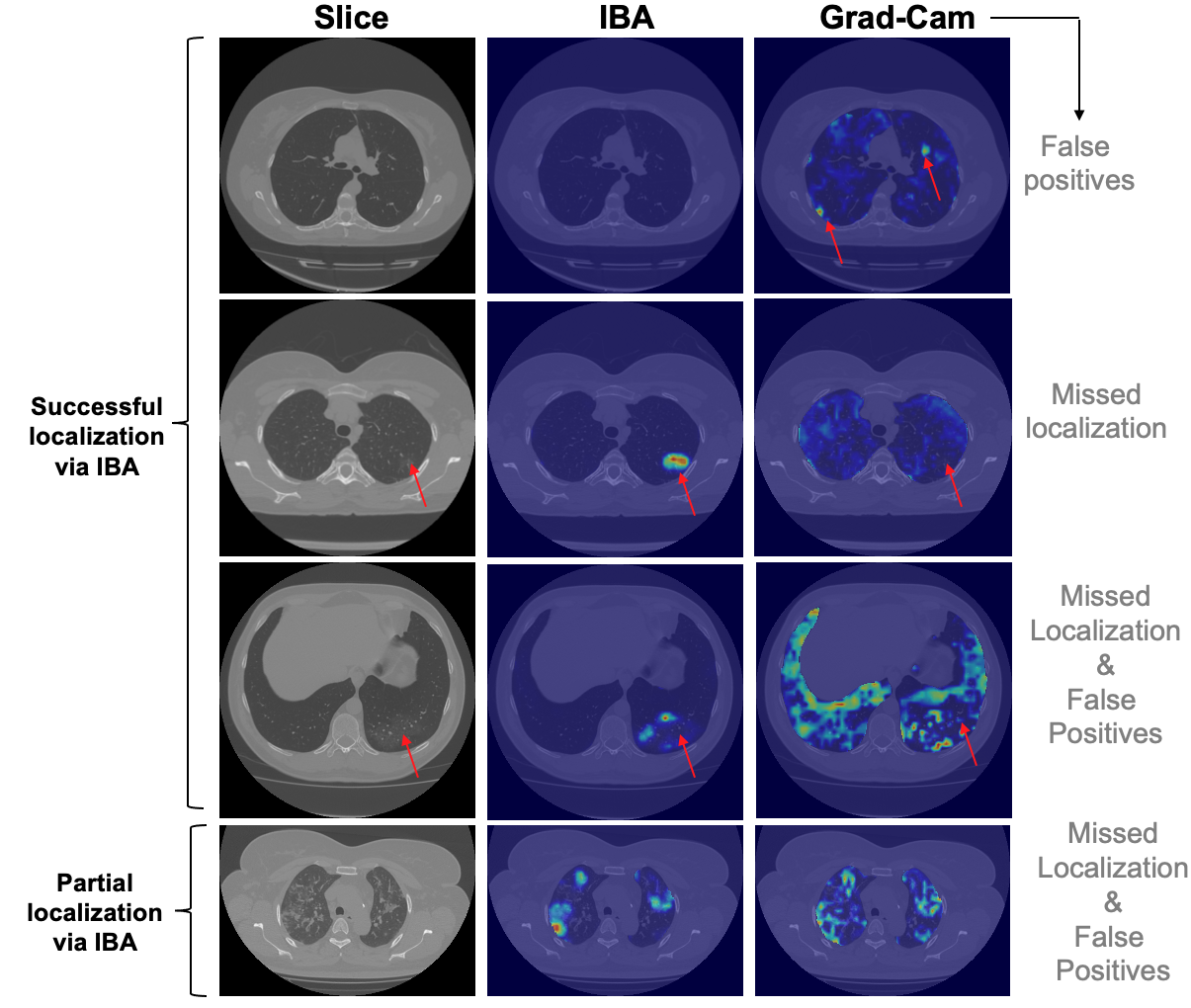}
\caption{Example heatmap predictions from IBA and Grad-CAM methods. In most of the examples IBA produces more precise localization results compared to Grad-CAM. Grad-CAM tends to over-predict localization.} \label{fig:heatmap_samples}
\end{figure}

\section{Experiments and Results}
Our pre-trained ResNeSt classifier achieved 73.87\% accuracy, 74.71\% sensitivity and 71.05\% specificity on the test set. This model was used both for IBA and Grad-CAM evaluations. As long as the same network architecture is used for backbone classification, comparisons of IBA and Grad-CAM will be fair. Any improvement in classification improvements will improve localization ability of both methods.

\textbf{Visual Evaluations} 
We have conducted two visual scoring experiments with four participating physicians who have never seen the data set that we have used in our study. We have selected representative CT scans from each severity class (30 CT scans) and fused both IBA and Grad-CAM visual results with original CT scans separately and used the same contrast/brightness windows for the visual inspection. In the first experiments, we asked four participating physicians to evaluate whether IBA or Grad-CAM is performing better, or tie. Figure~\ref{fig:vis} (left) shows visual inspection results for this part of the experiments: IBA wins these experiments with a large margin, and found to be equal to Grad-CAM only in a few cases. In the majority of the cases, IBA was found to be most successful. In the second experiment, we asked our participating physicians to evaluate the localization performance of the IBA method with the following scoring system: (1) satisfactory localization, (2) partial localization, (3) poor localization, and (4) no localization/unacceptable. Figure~\ref{fig:vis} (right) summarizes the visual scoring for this experiment: the majority of the cases were found to be successfully locating pathology regions with the IBA method, and only a few cases were found to be either partially or poorly localized.

\textbf{Qualitative Results} In Figure \ref{fig:heatmap_samples}, we demonstrated sample outputs from the IBA and Grad-CAM approaches. IBA finds sharp and finer regions, mostly tries to focus on only the necessary lesion as we expected. On the other hand, Grad-CAM overestimates the focused regions, and there are a large number of false-positive findings. Interestingly, even for negative findings (first row in Figure \ref{fig:heatmap_samples}) are considered to have Covid-19.  Since Grad-CAM does not have any mechanism to enforce networks to use information from a certain region, these results are not surprising. Another drawback of the Grad-CAM is that it is sensitive to hyper-parameters due to the gradient operation. Figure \ref{fig:heatmap_samples} exemplifies failure cases of Grad-CAM in varying levels of Covid-19 severity and shows the superiority of IBA. In the majority of the CT scans, as mentioned earlier, there is satisfactory localization with IBA, and only with a few cases, IBA was found to be partially overlapping. 

\begin{figure}
\centering
\includegraphics[width=1\textwidth]{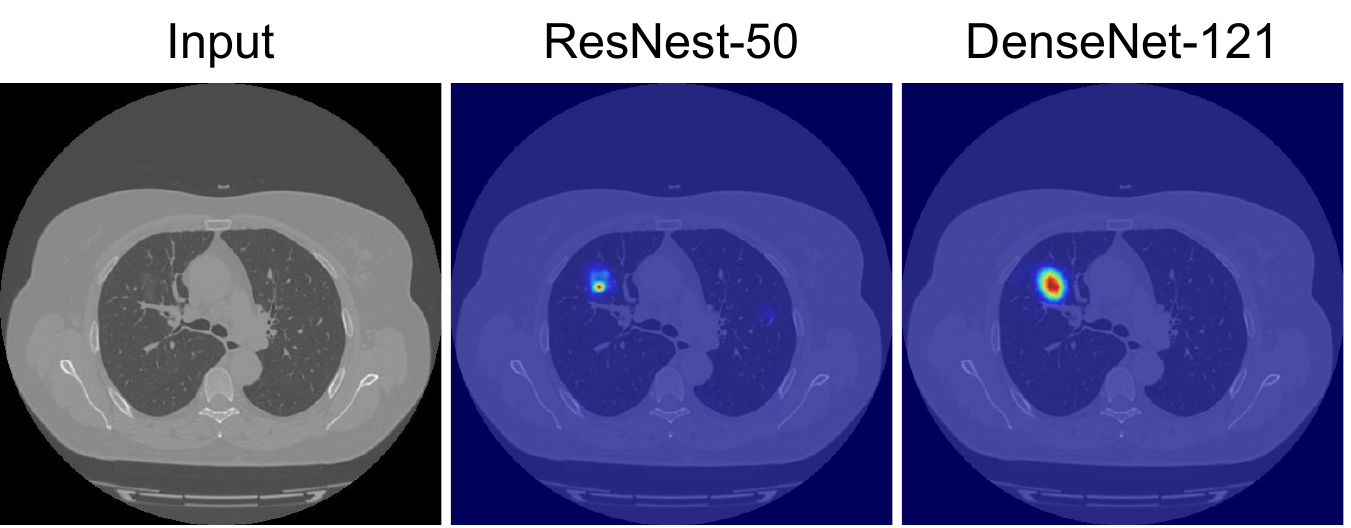}
\caption{IBA results from different network architectures.} \label{fig:arch_results}
\end{figure}

To show the effectiveness of IBA approach, alternatively, we used also a different network architecture, DenseNet, as a classifier model. Figure \ref{fig:arch_results} shows IBA results from ResNeSt and DenseNet architectures. In both examples, IBA locates similar regions on the input image. 


\begin{figure}
\centering
\includegraphics[width=\textwidth]{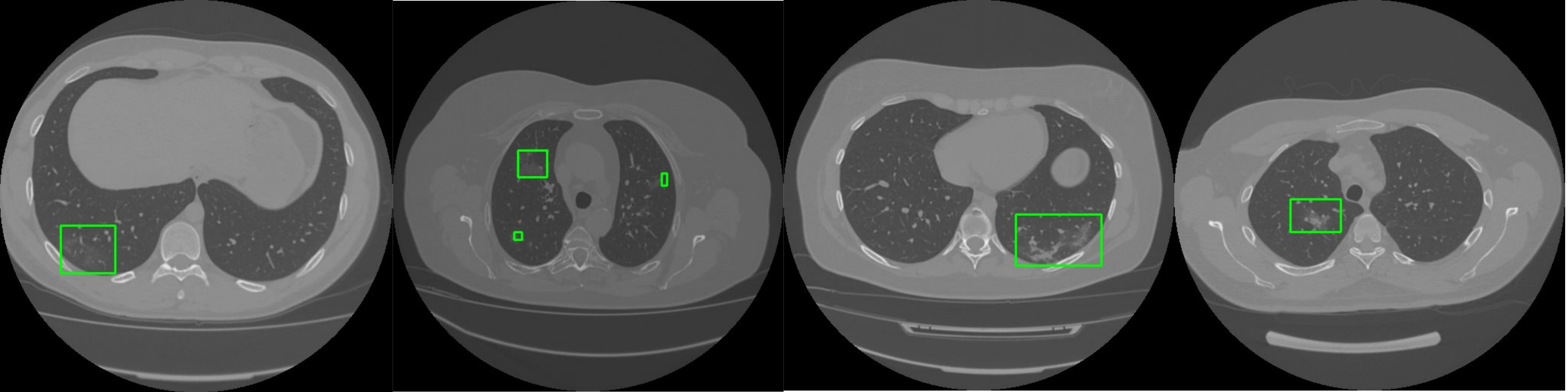}
\caption{Lesion localization by using IBA heatmaps.} \label{fig:bbox}
\end{figure}

\textbf{Lesion Detection} After obtaining the IBA heatmaps, we applied connected component analysis to detect the location of lesions. Figure \ref{fig:bbox} demonstrates the detection results for different inputs. These results show that lesions of varying sizes and multiple locations can be successfully located from IBA heatmaps.

\section{Discussion and Concluding Remarks}
This study sets out to critically examine the visual explanation methods for medical imaging. Our findings indicate that the information bottleneck-based approach, IBA, outperforms the Grad-CAM method both quantitatively and qualitatively..
It should be noted that our proposed method has certain limitations that we plan to address in our extended study. For instance, IBA is highly sensitive in localizing the Covid-19 related pathologies; however, IBA tends to underestimate the region of interest contributing to the diagnosis and prognosis. This is understandable because information is minimized to increase most relevant features to be captured. Similar equations can be set up for background identification to handle this discrepancy. The analysis of information bottleneck undertaken here has extended our knowledge about more robust localization methods that can have a vital effect in emergency situations such as the Covid-19 pandemic where we need to develop automated image analysis tools without having dense annotations. While we choose an application with an emerging need globally, IBA is general method that we plan to demonstrate its effectiveness in various different medical diagnosis tasks in our  future work. 

%
%
%
 \bibliographystyle{splncs04}
 \bibliography{mybibliography}

\end{document}